\documentclass[twocolumn,trackchanges]{aastex701}

\usepackage{newtxtext,newtxmath}
\usepackage[T1]{fontenc}
\usepackage{ae,aecompl}

\usepackage{graphicx}	% Including figure files
\usepackage{amsmath}	% Advanced maths commands
\usepackage{amssymb}	% Extra maths symbols

\usepackage{newtxtext,newtxmath}
\begin{document}

\title{Lack of Significant Orbital-Phase Locking in the Active Phases of the Recurrent Nova T CrB}

\author[orcid=0000-0002-0851-8045,sname='Pei']{Songpeng Pei}
\affiliation{School of Physics and Electrical Engineering, Liupanshui Normal University, Liupanshui, Guizhou, 553004, China}
\email[show]{songpengpei@outlook.com}

\author[0009-0004-5610-6549,sname=Zhang,gname=Xiaowan]{Xiaowan Zhang}
\affiliation{School of Physics and Electrical Engineering, Liupanshui Normal University, Liupanshui, Guizhou, 553004, China}
\email{xiaowanzhang001@outlook.com}

\author[0000-0002-2432-2587,sname=Su,gname=Renzhi]{Renzhi Su}
\affiliation{Shanghai Astronomical Observatory, Chinese Academy of Sciences, 80 Nandan Road, Shanghai 200030, China}
\email{rzsu.astro@gmail.com}

\author[0000-0002-7714-493X,sname=Cai,gname=Yongzhi]{Yongzhi Cai}
\affiliation{INAF - Osservatorio Astronomico di Padova, vicolo dell'Osservatorio 5, I-35122 Padova, Italy}
\affiliation{Yunnan Observatories, Chinese Academy of Sciences, Kunming 650216, China}
\affiliation{International Centre of Supernovae, Yunnan Key Laboratory, Kunming 650216, China}
\email{yzcai789@163.com}

\author[0000-0002-3632-474X,sname=Ou,gname=Ziwei]{Ziwei Ou}
\affiliation{Tsung-Dao Lee Institute, Shanghai Jiao Tong University, Shanghai 201210, China}
\email{ziwei@sjtu.edu.cn}

\author[0000-0002-7083-8007,sname=Li,gname=Qiang]{Qiang Li}
\affiliation{School of Physics and Electronic Science, Qiannan Normal University for Nationalities, Duyun 558000, China}
\affiliation{Qiannan Key Laboratory of Radio Astronomy, Guizhou Province, Duyun 558000, China}
\email{nliqiang@foxmail.com}

\author[sname=Ren,gname=Xiaoqin]{Xiaoqin Ren}
\affiliation{School of Physics and Electronic Science, Qiannan Normal University for Nationalities, Duyun 558000, China}
\affiliation{Qiannan Key Laboratory of Radio Astronomy, Guizhou Province, Duyun 558000, China}
\email{18786619324@163.com}

\author[0000-0002-4421-7282,sname=Liu,gname=Yu]{Yu Liu}
\affiliation{State Key Laboratory of Public Big Data, Guizhou University, Guiyang 550025, China}
\email{yuliu@gzu.edu.cn}

\author[0000-0002-1859-4949,sname=Yang,gname=Taozhi]{Taozhi Yang}
\affiliation{Ministry of Education Key Laboratory for Nonequilibrium Synthesis and Modulation of Condensed Matter, School of Physics, Xi'an Jiaotong University, 710049 Xi'an, China}
\email{yangtaozhi2018@163.com}

\begin{abstract}
T Coronae Borealis (T CrB) is a symbiotic recurrent nova (RN) that exhibits both nova eruptions and long-term active phases resembling superoutbursts and normal outbursts. Motivated by proposed connections between these events and the binary orbit, we test whether the onset, maximum, or termination of the active phases is locked to orbital phase. We use long-term optical $B$- and $V$-band light curves from the American Association of Variable Stars Observers (AAVSO) International Database and historical photometry from the literature. We measure the onset, maximum, and termination times of superoutbursts and normal outbursts and convert these times to orbital phase. We test the resulting circular distributions with Kuiper and Watson statistics. We find no statistically significant orbital-phase locking. The onset phases and maxima are consistent with a uniform phase distribution. The smallest probabilities occur for the 13 measurable termination phases ($p_{\rm MC}=0.043$ for the Kuiper statistic and $p_{\rm MC}=0.048$ for the Watson statistic), but this result is only marginal in an uncorrected $p<0.05$ sense, far from a $3\sigma$ detection, and insufficient to establish robust phase locking. The four historical nova eruptions likewise do not provide robust evidence for a unique ignition phase once the small sample size, historical date uncertainties, and long-term period changes are considered. The two known secondary eruptions occurred at similar phases, but two events are insufficient to establish an orbital-geometry connection. Overall, the active phases of T CrB appear to be governed primarily by accretion-disk physics rather than by a fixed binary phase.
\end{abstract}



\section{Introduction}\label{sec:intro}

T CrB (HR 5958; HD 143454) is the nearest known RN and one of the brightest and best-studied systems of its class. Its recorded eruptions in 1866 and 1946 both reached approximately $V \approx 2$~mag \citep{1949ApJ...109...81S}, and additional historical eruptions have been proposed for AD 1217 and 1787 \citep{2023JHA....54..436S, 2023ATel16107....1S}. Because nearly eight decades have elapsed since the 1946 eruption, T CrB has been widely regarded as being close to its next nova event \citep{2016NewA...47....7M, 2020ApJ...902L..14L, 2023MNRAS.524.3146S, 2023ApJ...953L...7I, 2023A&A...680L..18Z, 2025MNRAS.541L..14M, 2026A&A...706A..94P}. Recent predictions have placed the forthcoming eruption broadly in the interval 2023.0--2026.8 \citep{2020ApJ...902L..14L, 2023ATel16107....1S, 2023MNRAS.524.3146S}.

T CrB is a symbiotic binary composed of a Roche-lobe--filling M4~III red giant \citep{1998MNRAS.296...77B, 1999A&AS..137..473M, 2025ApJ...983...76H} and a very massive white dwarf (WD; $M_{\rm WD}=1.37\pm0.01\,M_\odot$; \citealt{2025ApJ...983...76H}). The orbital period is close to 227.57--227.58~d \citep{2000AJ....119.1375F, 2023MNRAS.524.3146S, 2025A&A...694A..85P}, with the recent circular radial-velocity solution giving $P_{\rm orb}=227.5494\pm0.0049$~d and $T_0={\rm HJD}\,2455427.51\pm0.10$ as the epoch of maximum positive velocity of the red giant \citep{2025ApJ...983...76H}. The eccentric solution of \citet{2025ApJ...983...76H} gives only $e=0.0072\pm0.0026$, and those authors favored the circular orbit. The optical light curve shows strong ellipsoidal modulation at half of the orbital period \citep{1975JBAA...85..217B, 2023MNRAS.524.3146S}. The system therefore combines properties of a symbiotic binary, a cataclysmic variable, and a near-Chandrasekhar-mass nova progenitor candidate. In addition to its nova eruptions, T CrB exhibits long-term active phases and a recent high state that has been interpreted as analogous to a superoutburst in an extreme SU~UMa-like system \citep{2023ApJ...953L...7I}. \citet{2023MNRAS.524.3146S} further emphasized the similarity between the historical light curves surrounding the 1866 and 1946 eruptions, including the pre-eruption high state, the pre-eruption dip, and the post-eruption secondary eruption.

Following the 1946 nova eruption, T CrB has undergone alternating phases of quiescence and enhanced activity. Around 2014--2015, it entered a high state similar to that observed before the 1946 eruption \citep{2016NewA...47....7M, 2016MNRAS.462.2695I, 2018A&A...619A..61L}. During this phase, the UV, optical, and radio emission increased markedly, the usual $B$-band orbital modulation disappeared, high-ionization lines strengthened, and the hard X-ray flux nearly vanished \citep{2016NewA...47....7M, 2018A&A...619A..61L, 2019ApJ...884....8L}. This behavior has been interpreted as a superoutburst-like state in an extreme SU~UMa-like dwarf nova \citep{2023ApJ...953L...7I} and as a possible precursor of the next nova eruption \citep{2016NewA...47....7M, 2020ApJ...902L..14L, 2023MNRAS.524.3146S, 2023ApJ...953L...7I, 2023A&A...680L..18Z}.

In symbiotic recurrent novae, the ignition mass on the WD is likely to be reached during an accretion high state \citep{2020ApJ...902L..14L}. For T CrB, it has therefore been argued that the WD accumulates most of the ignition mass during the high state \citep{2020ApJ...902L..14L, 2023A&A...680L..18Z, 2025A&A...701A.176M}. Compared with the enhanced mass-transfer phase preceding the 1946 eruption, the present high state attained a lower brightness, suggesting that a smaller amount of mass has been transferred from the disk to the WD \citep{2025A&A...701A.176M}. This may help explain why the next nova eruption had not occurred two full years after the end of the recent enhanced mass-transfer phase, whereas the 1946 nova followed the preceding high state after only about six months \citep{2025A&A...701A.176M}. In this context, the next nova eruption of T CrB is still expected to occur after the high state. We also note that the 2014--2023 high state corresponds to the superoutburst with the largest $B$-band amplitude among the post-1946 superoutbursts identified so far \citep{2023ApJ...953L...7I}.

The physical expectation for strict orbital-phase locking is not strong. In thermonuclear novae, ignition occurs in the high-pressure layer on the WD surface, while in dwarf-nova-like active phases the onset of thermal-viscous instability is expected to depend mainly on the mass, radius, and thermal state of the accretion disk rather than on the instantaneous binary phase. T CrB is not viewed face-on, with published inclination estimates in the range $i\simeq46$--$70^\circ$ \citep{1998MNRAS.296...77B, 2004A&A...415..609S, 2025ApJ...983...76H, 2025A&A...701A.176M}, so orbital geometry is not irrelevant; however, any direct phase dependence is expected to be weak. Moreover, because the orbit is essentially circular, there is no well-defined periastron passage at which one would naturally expect periodically enhanced Roche-lobe overflow or wind accretion. Similarly, ordinary dwarf novae and SU~UMa-type systems show statistical relations between orbital period and global outburst properties, but there is no generally established requirement that the onset, maximum, or termination of individual outbursts occur at a fixed binary orbital phase \citep{1989PASJ...41.1005O, 2005PJAB...81..291O, 2013PASJ...65...50O, 2016MNRAS.460.2526O}. Thus, any orbital-phase preference in T CrB should be treated as an empirical hypothesis to be tested rather than as an expected property of disk outbursts.

Despite this weak theoretical expectation, several empirical facts motivate a direct test. \citet{2023ApJ...953L...7I} argued that the active phases are consistent with normal outbursts and superoutbursts in an SU~UMa-like dwarf nova, whereas \citet{2024RNAAS...8..272S} pointed out that the historical nova eruptions are separated by approximately integer multiples of the orbital period, suggesting a possible empirical connection between eruption timing and orbital phase. At the same time, \citet{2023MNRAS.524.3146S} showed that the onset of the secondary eruption after the 1866 and 1946 novae occurred at nearly the same post-eruption stage, raising the possibility that orbital geometry may influence some parts of the outburst phenomenology more strongly than others. 

Additional phenomenological hints also point to a possible connection between long-term variability and the orbital period. During the plateau of the most recent big active phase, variability on a timescale of $\sim 700$~d is visible \citep{2023ApJ...953L...7I}. In particular, the interval between the first and second peaks is $\sim 680$~d, while that between the third and fourth peaks is $\sim 690$~d (see Fig.~4 in \citealt{2023ApJ...953L...7I}); these values correspond to 3.076, 2.988, and 3.032 times the orbital period of 227.5687~d, respectively, adopting the ephemeris of \citet{2000AJ....119.1375F}. Likewise, the minimum recurrence times of normal outbursts and superoutbursts used by \citet{2023ApJ...953L...7I}, namely $P_{\mathrm{c}}=906$~d and $P_{\mathrm{sc}}=3640$~d from \citet{1997ppsb.conf..117A}, correspond to 3.981 and 15.995 times the orbital period, respectively. These near-integer relations are quoted here in the same Fekel-ephemeris convention used by \citet{2024RNAAS...8..272S}; in the phase analysis below, however, we adopt the \citet{2025ApJ...983...76H} ephemeris. These near-integer relations do not demonstrate phase locking, but they justify testing whether the onset, peak, or termination of active phases is preferentially distributed in orbital phase.

This work is therefore framed as a falsification test for strong orbital-phase locking, not as a claim that such locking is theoretically expected. Our data set and analysis also differ from the main earlier studies. \citet{2023MNRAS.524.3146S} constructed the long-term $B$- and $V$-band light curves and studied the high states, eruption morphology, and orbital period changes. \citet{2023ApJ...953L...7I} interpreted the active phases as normal outbursts and superoutbursts and measured several maxima. Here we use the processed long-term light curve to measure onset, maximum, and termination times for four superoutbursts and ten normal outbursts, convert all three characteristic times to orbital phase, and evaluate the phase distributions with circular statistics. We also discuss the historical nova eruptions and the two known secondary eruptions, but only as small-number consistency checks.

\section{Observations and Data}\label{sec:observation}

We used long-term optical photometry of T CrB in the $B$ and $V$ bands. The core of the data set was taken from the AAVSO International Database (AID)\footnote{\url{https://www.aavso.org/data-access} \citep{Kloppenborg2025AAVSO}}, from which we retrieved publicly available observations of T CrB in these filters. Following \citet{2023MNRAS.524.3146S}, the $B$- and $V$-band light curves from the AAVSO International Database were binned to 0.01~yr intervals. To extend the historical baseline and improve the coverage of earlier activity, we also incorporated the optical measurements compiled by \citet{2023MNRAS.524.3146S} and the references assembled therein. This combined data set provides broad temporal coverage of the nineteenth- and twentieth-century nova events, the long inter-eruption interval, and the recent high state.

\begin{figure*}
\centering
\includegraphics[width=\textwidth]{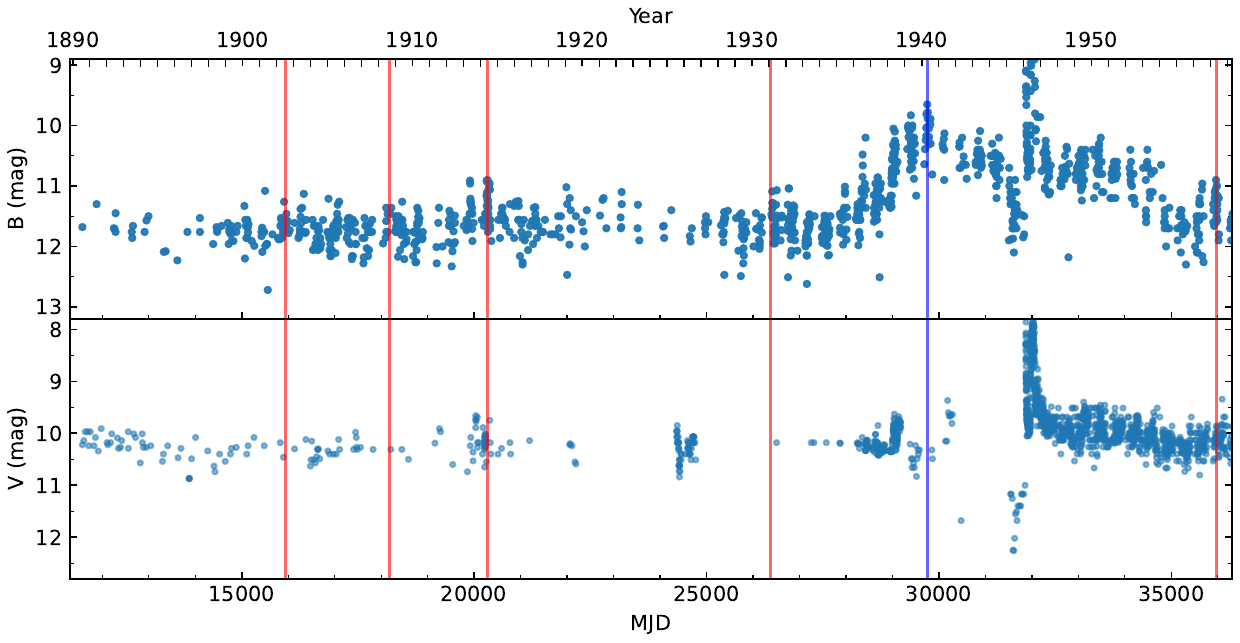}
\vspace{2mm}
\includegraphics[width=\textwidth]{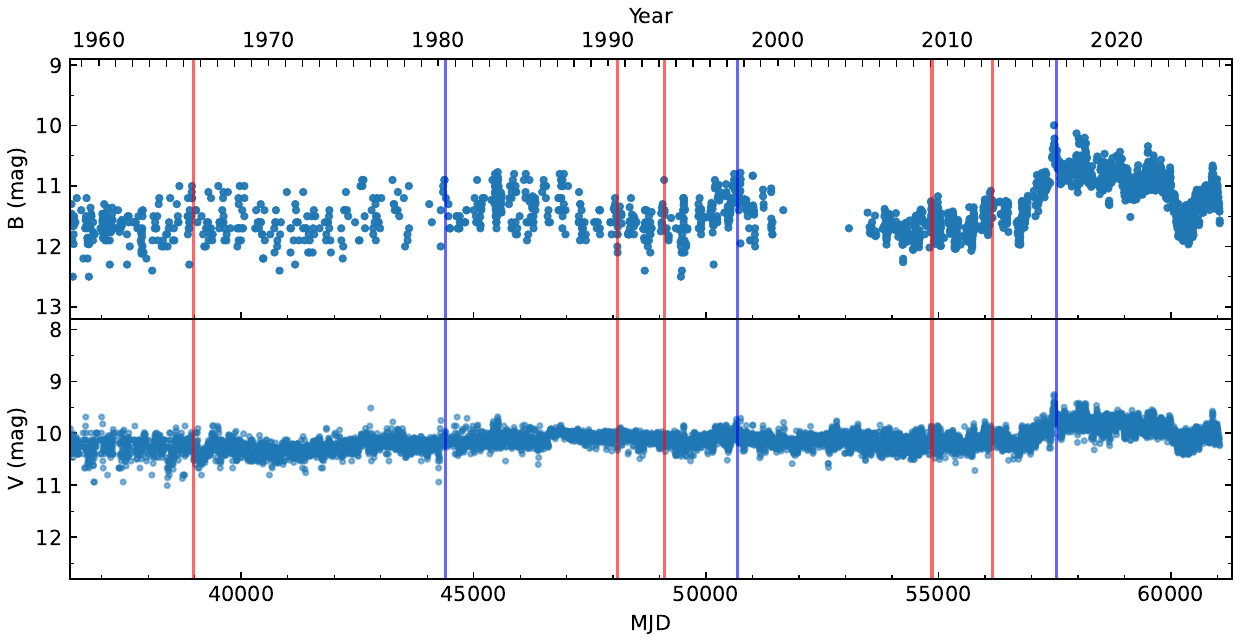}
\caption{Historical optical light curves of T CrB used to identify the active phases analyzed in this work. The upper panel shows the earlier interval, MJD 11571.0--36311.0, and the lower panel shows the later interval, MJD 36311.0--61051.0. In each panel, the $B$-band light curve is shown above the $V$-band light curve. Blue vertical lines mark the maxima of superoutbursts and red vertical lines mark the maxima of normal outbursts. The full active-phase durations are shown in the processed-light-curve version of the figure below.}
\label{fig:lc_hist}
\end{figure*}

The historical light curves used in this work are shown in Figure~\ref{fig:lc_hist}. These data include the major active phases identified in previous studies and the additional events recognized in the present analysis. Because our goal is to test the orbital-phase distribution of event boundaries rather than to model short-timescale stochastic variability, this optical data set is well suited to the purpose: it spans multiple decades, samples both quiescent and active states, and preserves the long-term morphology required to identify superoutbursts, normal outbursts, and their characteristic times.

Further processing steps, including subtraction of the orbital ellipsoidal modulation, LOESS smoothing, removal of the red-giant contribution, and measurement of onset, maximum, and termination times, are described in Section~\ref{sec:phase}.

\section{Orbital-Phase Analysis}\label{sec:phase}

To test whether the onset, maximum, and termination of active phases in T CrB occur preferentially at specific orbital phases, we measured these characteristic times from the long-term optical light curve and mapped them onto the binary orbit using a fixed spectroscopic ephemeris.

\subsection{Smoothed Light Curve of the Active Phases}\label{subsec:smoothed}

We determined the times of maxima of the big active phases (superoutbursts) and small active phases (normal outbursts) in the $B$ band following the general procedure adopted by \citet{2023ApJ...953L...7I}. The long-term $B$- and $V$-band light curves were first corrected for orbital ellipsoidal modulation by subtracting sinusoidal components with amplitudes of $0.173(5)$~mag in $V$ and $0.172(12)$~mag in $B$, adopting these values directly from \citet{2023ApJ...953L...7I} rather than re-deriving them.

The orbital-cleaned light curves were subsequently smoothed using LOESS. LOESS is a local polynomial regression method in which the value of the smooth curve at each epoch is determined from nearby data, weighted here by a Gaussian kernel. We used the \texttt{localreg} package\footnote{\url{https://zenodo.org/records/6344451}} (v.~0.5.0; \citealt{2022zndo...6344451M}) and adopted a characteristic radius of 113.71~d, as in \citet{2023ApJ...953L...7I}. This smoothing radius is close to half of the binary period, but here it is used only as a low-pass filter for the long-term light curve, not as an imposed orbital clock. Because some active phases last several orbital cycles, changing the smoothing prescription can move the formal crossing times by a non-negligible fraction of an orbit. We therefore use the resulting dates as operational descriptors of each active phase, not as uniquely defined physical trigger times. This smoothing suppresses short-timescale flickering and emphasizes the secular evolution of the active phases.

\begin{figure*}
\centering
\includegraphics[width=\textwidth]{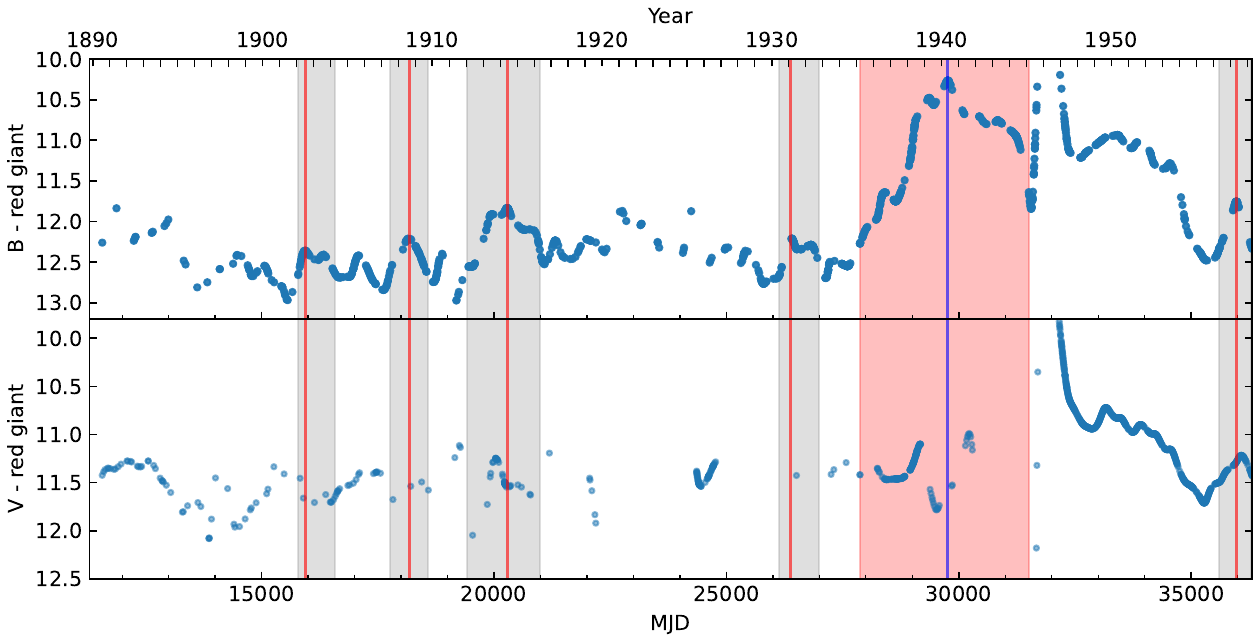}
\vspace{2mm}
\includegraphics[width=\textwidth]{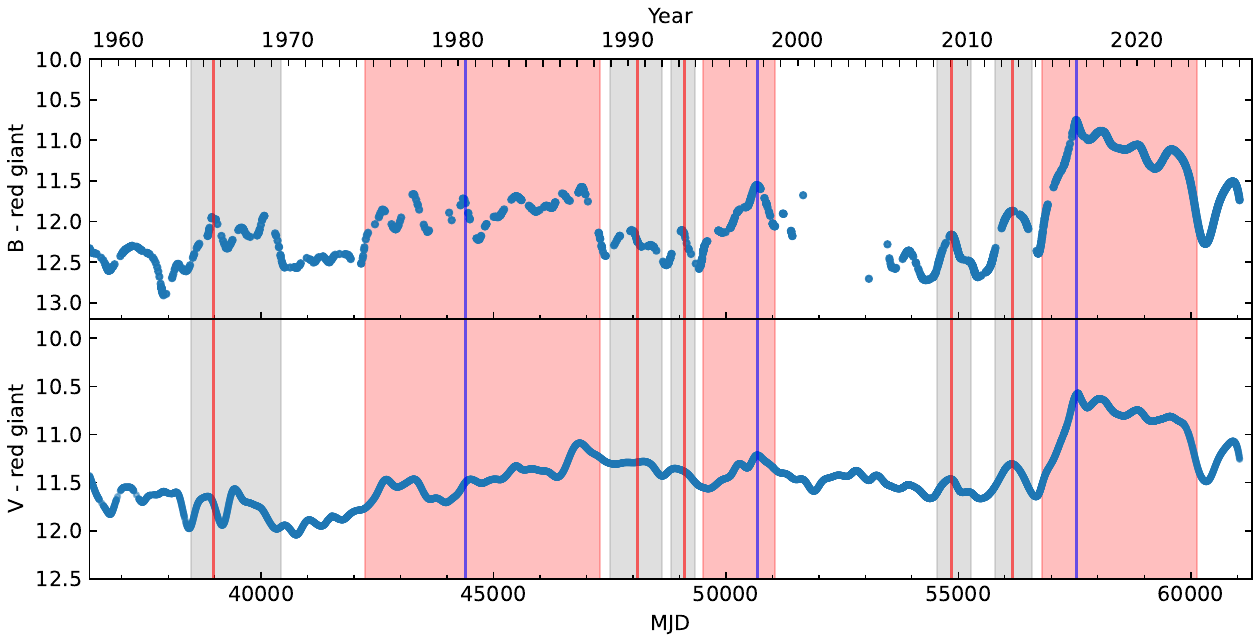}
\caption{Processed optical light curves of T CrB after subtraction of the orbital ellipsoidal modulation, LOESS smoothing, and removal of the red-giant contribution in flux space. The upper panel shows MJD 11571.0--36311.0 and the lower panel shows MJD 36311.0--61051.0. In each panel, the $B$-band light curve is shown above the $V$-band light curve. Semi-transparent shaded regions indicate the full extent of each active phase, while the vertical lines mark the adopted maxima. This visual distinction emphasizes that the active stages are extended in time and that the single maximum time is only one operational descriptor of each broad event. These processed curves were used to identify active phases and measure the characteristic times listed in Table~\ref{tab:times}.}
\label{fig:lc_processed}
\end{figure*}

We then removed the red-giant contribution in flux space. For this purpose, we adopted $V_{\rm RG}=10.029$~mag \citep{2004MNRAS.350.1477Z}, $E(B-V)=0.15$~mag \citep{1992ApJ...393..289S}, and $(B-V)_{\rm RG}=1.60$~mag \citep{1992msp..book.....S}, appropriate for an M4.5~III giant \citep{1999A&AS..137..473M}. The observed magnitudes and the theoretical red-giant magnitudes were converted to fluxes, the red-giant contribution was subtracted, and the residual fluxes were converted back to magnitudes. The resulting smoothed, orbital-cleaned, and red-giant-subtracted light curves are shown in Figure~\ref{fig:lc_processed}.

The final processed $B$-band light curve was used to identify active phases and measure their characteristic times. Relative to the sample discussed by \citet{2023ApJ...953L...7I}, we include one additional superoutburst and six additional normal outbursts in the historical record. The $B$ band was adopted for the timing measurements because it provides the best combination of cadence, historical coverage, and accretion-disk contrast. We do not assume that the $V$-band maximum or boundary time must be identical to the $B$-band value; where the $V$ band is sufficiently sampled, chromatic offsets are possible and represent an additional source of timing uncertainty. We did not repeat the full boundary-measurement procedure in the $V$ band because the brightness changes during the active phases are weaker in $V$ than in $B$, making the broad onset and termination crossings less robust. For this reason, our phase-locking test should be read as a uniform, $B$-band operational test of the active phases rather than a claim that all photometric bands define identical event times. The times of maxima are listed in Table~\ref{tab:times}.

\subsection{Definition of Onset, Maximum, and Termination Times}\label{subsec:definitions}

For each active phase, following \citet{2023ApJ...953L...7I}, the time of maximum, $t_{\rm max}$, was defined as the epoch of peak brightness in the smoothed $B$-band light curve, including plateau-like maxima for superoutbursts and sharp local maxima for normal outbursts. The onset and termination times were defined relative to local pre-event and post-event baselines.

For a given event, let $m_{\rm max}$ be the magnitude at maximum, and let $m_{\rm base,pre}$ and $m_{\rm base,post}$ denote the local quiescent baseline magnitudes estimated on the pre-maximum and post-maximum sides, respectively. We define the pre-event and post-event amplitudes as
\begin{equation}
A_{\rm pre}=m_{\rm base,pre}-m_{\rm max},
\end{equation}
and
\begin{equation}
A_{\rm post}=m_{\rm base,post}-m_{\rm max}.
\end{equation}
We defined the onset and termination times on the peak-anchored, LOESS-smoothed, red-giant--subtracted $B$-band light curve as the times at which the curve crossed 15\% of the local event amplitude within restricted windows around each visually identified event boundary. We also tested 10\% and 20\% thresholds and found that the resulting event ordering and classification were unchanged. In magnitude space, the adopted 15\% thresholds are
\begin{equation}
m_{\rm th,pre}=m_{\rm base,pre}-0.15A_{\rm pre},
\end{equation}
and
\begin{equation}
m_{\rm th,post}=m_{\rm base,post}-0.15A_{\rm post}.
\end{equation}
The onset time, $t_{\rm on}$, was defined as the last epoch before $t_{\rm max}$ at which the smoothed light curve crossed $m_{\rm th,pre}$ in the brightening direction, while the termination time, $t_{\rm term}$, was defined as the first epoch after $t_{\rm max}$ at which the smoothed light curve crossed $m_{\rm th,post}$ in the fading direction. In practice, the search for each crossing was restricted to a local time window centered on the visually identified event boundary, in order to avoid contamination by neighboring activity and to keep the measurements tied to the individual outburst morphology. For the superoutburst peaking at MJD~50670, the post-maximum $B$-band coverage is insufficient to define a reliable 15\% termination crossing; we therefore leave its termination time blank in Table~\ref{tab:times} and exclude it from the termination-phase statistical test. The full onset-to-termination intervals for the 13 events with measurable terminations are shown as shaded regions in Figure~\ref{fig:lc_processed}. These broad intervals illustrate why a single characteristic date is an imperfect representation of an extended active stage. The statistical test below is therefore deliberately conservative: it asks whether these consistently defined operational times show robust phase clustering. The resulting onset, maximum, and termination times are listed in Table~\ref{tab:times}.

\subsection{Orbital Phases and Phase Uncertainties}\label{subsec:phases}

Because \citet{2024RNAAS...8..272S} used the \citet{2000AJ....119.1375F} spectroscopic ephemeris, that ephemeris remains useful for discussing the published integer-cycle relation. For the quantitative phase analysis in the present work, however, we adopt the circular spectroscopic solution of \citet{2025ApJ...983...76H}. This solution combines the earlier radial velocities, the \citet{2000AJ....119.1375F} data, the \citet{2025A&A...694A..85P} velocities, and new Automatic Spectroscopic Telescope (AST) velocities, and it is therefore preferable for the present phase calculation. The \citet{2025A&A...694A..85P} period, $P=227.58\pm0.03$~d, is consistent with the adopted value but is less precise and uses a different phase-zero convention.

\begin{figure*}
\centering
\includegraphics[width=\textwidth]{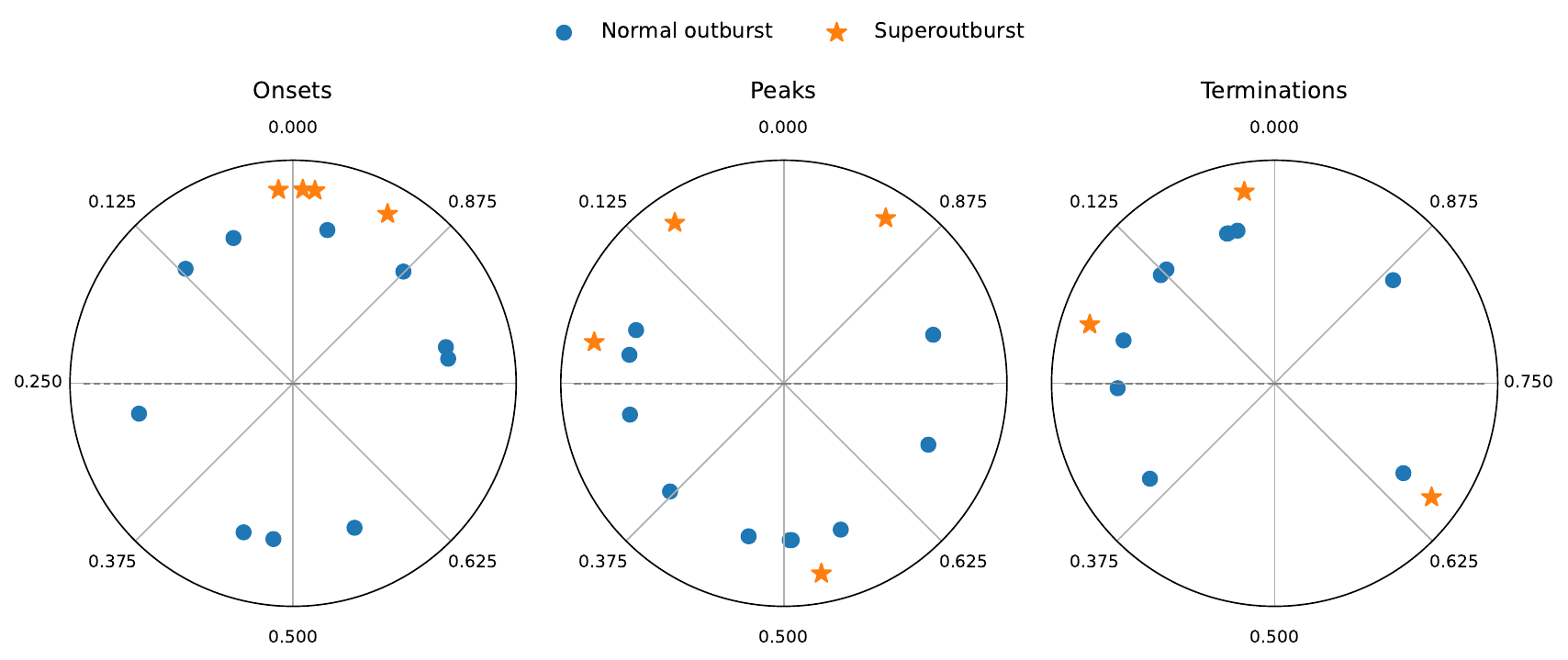}
\caption{Polar representation of the orbital-phase distributions of the onset times (left), maxima times (middle), and termination times (right) of the active phases in T CrB. Orbital phases were computed using the circular spectroscopic ephemeris of \citet{2025ApJ...983...76H}. Blue symbols denote normal outbursts and orange symbols denote superoutbursts. The visual distribution is consistent with the statistical tests in Table~\ref{tab:stats}: the onset and maximum samples show no significant clustering, while the termination sample shows only a marginal hint rather than a robust detection of orbital-phase locking.}
\label{fig:phase_polar}
\end{figure*}

Orbital phases were therefore computed with
\begin{equation}
P_{\rm orb}=227.5494~{\rm d}, \qquad T_0={\rm HJD}\,2455427.51,
\end{equation}
where $T_0$ corresponds to the epoch of maximum positive radial velocity of the red giant in the circular orbit of \citet{2025ApJ...983...76H}. For an event time $t$ in MJD, we used
\begin{equation}
\phi(t)=\left[\frac{t+2400000.5-T_0}{P_{\rm orb}}\right] \bmod 1.
\end{equation}
The orbital phases of the onset, maximum, and termination times are listed in Table~\ref{tab:times} and displayed in Figure~\ref{fig:phase_polar}.

The formal phase uncertainty from the ephemeris may be written approximately as
\begin{equation}
\sigma_\phi \simeq
\left[
\left(\frac{\sigma_t}{P_{\rm orb}}\right)^2+
\left(\frac{\sigma_{T_0}}{P_{\rm orb}}\right)^2+
\left(\frac{(t+2400000.5-T_0)\sigma_P}{P_{\rm orb}^2}\right)^2
\right]^{1/2}.
\end{equation}
where $\sigma_t$ is the uncertainty in the measured event time, $\sigma_{T_0}$ and $\sigma_P$ are the uncertainties of the reference epoch and orbital period, respectively. This expression is the standard first-order propagation of uncertainty for the linear ephemeris used in Equation~(6), before applying the modulo operation. For the active phases, the dominant uncertainty is not the formal spectroscopic ephemeris, but the identification of broad onset and termination boundaries in a smoothed long-term light curve. Such uncertainties would smear any real clustering rather than create a false one. We therefore treat the tabulated active-phase phases as point estimates for the statistical tests, while interpreting any marginal clustering conservatively. For the post-1946 active phases, replacing the adopted \citet{2025ApJ...983...76H} ephemeris with the \citet{2000AJ....119.1375F} ephemeris shifts the tabulated phases by only $-0.011$ to $+0.006$ cycles. The statistical conclusion is unchanged. The difference is much larger for the medieval and eighteenth-century nova dates because of the much longer extrapolation baseline, which reinforces the caution applied to the historical nova phases.

The situation is different for the historical nova eruptions. The dates of the 1217 and 1787 events are intrinsically uncertain, and \citet{2023MNRAS.524.3146S} found evidence that the orbital period of T CrB is not constant over the full historical baseline. Consequently, the phases of the historical nova eruptions computed with any fixed modern ephemeris are useful for comparison with previous work, but they should not be interpreted as precise geometric phases for the medieval and eighteenth-century events. This limitation applies even more strongly when comparing the \citet{2000AJ....119.1375F} and \citet{2025ApJ...983...76H} ephemerides over the full 1217--1946 interval.

\subsection{Circular Statistical Tests}\label{subsec:statistics}

Our treatment of phase distributions follows standard methods for circular data analysis \citep{Jammalamadaka2001}. For a sample of $N$ phases sorted as $0\leq x_1\leq x_2\leq \cdots \leq x_N<1$, the Kuiper statistic is
\begin{equation}
V=D^+ + D^-,
\end{equation}
with
\begin{equation}
D^+=\max_i\left(\frac{i}{N}-x_i\right), \qquad
D^-=\max_i\left(x_i-\frac{i-1}{N}\right).
\end{equation}
The Kuiper test is a circular analogue of the Kolmogorov-Smirnov test and is sensitive to localized deviations from uniformity while being invariant under a shift of phase origin \citep{Kuiper1960}.

We also used the Watson statistic,
\begin{equation}
U^2=\sum_{i=1}^N \left[x_i-\frac{2i-1}{2N}-\bar{d}\right]^2 + \frac{1}{12N},
\end{equation}
where
\begin{equation}
\bar{d}=\frac{1}{N}\sum_{i=1}^N \left[x_i-\frac{2i-1}{2N}\right].
\end{equation}
The Watson statistic measures the squared deviation of the empirical circular distribution from uniformity after removing the mean offset \citep{Watson1961}. For both statistics, large values indicate a stronger departure from uniformity.

Because the number of events is small, we calibrated significance with Monte Carlo simulations under the null hypothesis of a uniform phase distribution \citep{Press2007}. For each observed sample size, we generated synthetic phase samples drawn uniformly from $[0,1)$, computed the same statistic for each synthetic sample, and defined $p_{\rm MC}$ as the fraction of simulations with a statistic at least as large as the observed value. Thus, small $p_{\rm MC}$ indicates that the observed phase distribution is unusually non-uniform relative to a uniform parent distribution. In this paper, we use $p<0.05$ as a conventional criterion for a suggestive result and $p=0.0027$ as the two-sided $3\sigma$ criterion. As shown below, the onset and maximum samples do not reach the conventional $p<0.05$ level, while the termination sample is only marginal at this level and remains far from the $3\sigma$ criterion.

As an additional descriptive diagnostic, we computed the first trigonometric moment of each phase sample. This is equivalent to fitting the phase dependence with a first-harmonic form,
\begin{equation}
p(\phi) \propto 1+A\cos\left[2\pi(\phi-\phi_0)\right],
\end{equation}
where $A$ is a descriptive modulation amplitude and $\phi_0$ is the phase of the first-harmonic maximum. We use this quantity only to summarize the direction of any weak asymmetry; because the representative tests do not establish a robust orbital-phase-locking detection, we do not interpret the fitted $A$ or $\phi_0$ as physical parameters.

\subsection{Orbital-Phase Distribution of the Active Phases}\label{subsec:active_phases}

The active-phase tests are summarized in Table~\ref{tab:stats}. The onset phases of all active phases are consistent with uniformity. For the full onset sample, $V=0.3696$ with $p_{\rm MC}=0.202$ and $U^2=0.1191$ with $p_{\rm MC}=0.190$. These values provide no evidence for a preferred onset phase. The normal-outburst onset subsample is likewise consistent with uniformity. The superoutburst-onset subsample is visually clustered near phase zero, but it contains only four events; we therefore treat this as an exploratory small-number feature rather than as a robust detection of orbital triggering.

The peak phases are also consistent with a uniform distribution. For all outbursts combined, we obtain $V=0.3505$ with $p_{\rm MC}=0.275$ and $U^2=0.0894$ with $p_{\rm MC}=0.345$. We therefore find no evidence that the maxima of active phases occur at a preferred orbital phase.

The termination phases give the smallest probabilities among the three primary representative tests. For the 13 active phases with measurable terminations, we find $V=0.4626$ with $p_{\rm MC}=0.043$ and $U^2=0.1856$ with $p_{\rm MC}=0.048$. This is marginal in an uncorrected $p<0.05$ sense, but it does not satisfy a two-sided $3\sigma$ criterion ($p=0.0027$) and should be interpreted cautiously. The result also does not survive a simple correction for the three primary timing definitions tested here.

Taken together, the onset, maximum, and termination times of the active phases in T CrB show no robust statistically significant orbital-phase locking. The onsets and maxima are consistent with uniformity, while the termination phases show only a marginal hint of clustering. The data are therefore most naturally interpreted in a scenario in which active-phase evolution is controlled primarily by intrinsic accretion-disk physics rather than by the instantaneous orbital geometry.

To assess the sensitivity of the active-phase timings to the adopted LOESS smoothing scale, we repeated the measurement procedure for characteristic radii of 50, 60, 70, 80, 90, 100, 110, 120, and 130~d, keeping the event list and the restricted boundary windows fixed. The onset and maximum samples remained consistent with uniformity over this range. The termination sample continued to give the smallest probabilities for some smoothing radii, but these probabilities remained marginal, did not survive a simple correction for the three timing definitions tested, and never approached the two-sided $3\sigma$ threshold. Thus, the conclusion that the active phases do not show robust orbital-phase locking is not driven by the specific choice of the 113.71~d smoothing radius.

\subsection{Historical Nova Eruptions and Secondary Eruptions}\label{subsec:nova_phases}

For completeness, we also examined the orbital phases of the four recorded nova eruptions. Following \citet{2024RNAAS...8..272S}, we adopted the historical eruption dates JD~2165874, 2374102, 2402734, and 2431861 for the events in 1217, 1787, 1866, and 1946, respectively. Using the same \citet{2025ApJ...983...76H} ephemeris, these eruptions occurred at phases $\phi_{\rm start,1217}=0.5136$, $\phi_{\rm start,1787}=0.6028$, $\phi_{\rm start,1866}=0.4305$, and $\phi_{\rm start,1946}=0.4334$. These values do not define one unique nova-ignition phase.

If the four phases are treated as exact point estimates under the fixed ephemeris, Kuiper and Watson tests give point-estimate probabilities of order a few percent. This is not a robust detection. The sample contains only four events, the 1217 and 1787 eruption dates are historically uncertain, and the long-term orbital period changes reported by \citet{2023MNRAS.524.3146S} make backward extrapolation of a fixed modern ephemeris uncertain. We therefore use the historical nova phases only as a descriptive comparison and do not regard them as evidence for a unique orbital ignition phase.

The secondary eruptions after the 1866 and 1946 novae are also considered only descriptively. From the $V$-band light curves, which provide denser coverage of the post-eruption interval, we estimate an onset time of $\sim{\rm JD}\,2402846.3$ for the 1866 secondary eruption. This is $\sim112$~d after the primary eruption start on JD~2402734 and corresponds to $\phi_{\rm sec,1866}\approx0.924$. For the 1946 event, we estimate an onset time of $\sim{\rm JD}\,2431970.7$, or $\sim110$~d after the primary eruption start on JD~2431861, corresponding to $\phi_{\rm sec,1946}\approx0.916$. The similarity is noteworthy, but it involves only two events. Moreover, if the secondary maximum is caused by irradiation of the red giant by the cooling white dwarf, as proposed by \citet{2023RNAAS...7..251M}, then similar secondary-maximum phases would naturally be tied to the geometry and timing of the preceding primary nova eruption. In that case, the secondary eruptions are not an independent test of orbital-phase locking. It is therefore insufficient to establish an orbital-geometry connection or a physical triggering mechanism for the secondary eruption.

\begin{deluxetable*}{lccccccc}
\tabletypesize{\scriptsize}
\tablewidth{0pt}
\tablecaption{Measured onset, maximum, and termination times of the active phases in T CrB, together with the corresponding orbital phases derived from the circular spectroscopic ephemeris of \citet{2025ApJ...983...76H}. Times are measured from the processed $B$-band light curve described in Section~\ref{subsec:smoothed}. ``Normal'' denotes a small active phase/normal outburst, and ``Super'' denotes a big active phase/superoutburst. The final column identifies the literature source of maxima adopted from earlier work; entries marked ``This work'' were measured in the present analysis.\label{tab:times}}
\tablehead{
\colhead{Type} & \colhead{$t_{\rm onset}$$^a$} & \colhead{$\phi_{\rm onset}$$^b$} & \colhead{$t_{\rm maxima}$$^c$} & \colhead{$\phi_{\rm maxima}$$^d$} & \colhead{$t_{\rm term}$$^e$} & \colhead{$\phi_{\rm term}$$^f$} & \colhead{Maxima. source} \\
\colhead{} & \colhead{(MJD)} & \colhead{} & \colhead{(MJD)} & \colhead{} & \colhead{(MJD)} & \colhead{} & \colhead{}
}
\startdata
Normal & 15785 & 0.787 & 15939 & 0.464 & 16574 & 0.255 & This work \\
Normal & 17756 & 0.449 & 18173 & 0.282 & 18575 & 0.048 & This work \\
Normal & 19423 & 0.775 & 20284 & 0.559 & 20988 & 0.653 & This work \\
Normal & 26137 & 0.281 & 26385 & 0.371 & 26992 & 0.038 & This work \\
Super & 27875 & 0.919 & 29763 & 0.216 & 31540 & 0.025 & This work \\
Normal & 35602 & 0.876 & 35973 & 0.506 & 36282 & 0.864 & This work \\
Normal & 38470 & 0.480 & 38972 & 0.686 & 40438 & 0.129 & This work \\
Super & 42225 & 0.982 & 44398 & 0.531 & 47281 & 0.201 & \citet{1992ApJ...393..289S} \\
Normal & 47490 & 0.120 & 48100 & 0.800 & 48628 & 0.121 & \citet{1997IBVS.4461....1Z} \\
Normal & 48820 & 0.965 & 49100 & 0.195 & 49330 & 0.206 & \citet{1997IBVS.4461....1Z} \\
Super & 49509 & 0.992 & 50670 & 0.095 & \ldots & \ldots & \citet{2023ApJ...953L...7I} \\
Normal & 54531 & 0.062 & 54860 & 0.508 & 55280 & 0.354 & \citet{2023ApJ...953L...7I} \\
Normal & 55783 & 0.564 & 56160 & 0.221 & 56576 & 0.049 & \citet{2023ApJ...953L...7I} \\
Super & 56795 & 0.012 & 57455 & 0.912 & 60126 & 0.650 & \citet{2023ApJ...953L...7I} \\
\enddata
\tablecomments{
$^a $: The time of onset in MJD. $^b $: The orbital phase of the time of onset. $^c $: The time of maxima in MJD. $^d$: The orbital phase of the time of maxima. $^e $: The time of termination in MJD. $^f $: The orbital phase of the time of termination.
}
\end{deluxetable*}

\begin{deluxetable*}{lccccc}
\tablecaption{Representative Circular-Statistics Results\label{tab:stats}}
\tablehead{
\colhead{Sample} & \colhead{$N$} & \colhead{Kuiper $V$} & \colhead{$p_{\rm MC}(V)$} & \colhead{Watson $U^2$} & \colhead{$p_{\rm MC}(U^2)$}
}
\startdata
All onsets & 14 & 0.3696 & 0.202 & 0.1191 & 0.190 \\
All maxima & 14 & 0.3505 & 0.275 & 0.0894 & 0.345 \\
All terminations & 13 & 0.4626 & 0.043 & 0.1856 & 0.048 \\
\enddata
\tablecomments{$p_{\rm MC}$ is the Monte Carlo probability that a uniform phase distribution with the same sample size would yield a statistic at least as large as the observed statistic. The termination sample contains 13 events because the superoutburst peaking at MJD~50670 has no reliably measurable 15\% termination time in the $B$ band. The termination probabilities are marginal under an uncorrected $p<0.05$ criterion, but they are far from the two-sided $3\sigma$ threshold $p=0.0027$ and do not survive a simple correction for the three primary timing definitions tested here.}
\end{deluxetable*}

\section{Discussion}\label{sec:discussion}

\subsection{No Significant Orbital-Phase Locking of Active Phases}\label{subsec:discussion_main}

The main result of this study is negative but physically useful: the active phases of T CrB do not show robust statistically significant orbital-phase locking. With the \citet{2025ApJ...983...76H} ephemeris, the full onset and maximum samples are consistent with uniformity, with $p_{\rm MC}=0.202$ and 0.190 for the onset Kuiper and Watson tests and $p_{\rm MC}=0.275$ and 0.345 for the corresponding maximum tests. The 13 measurable termination phases give marginal uncorrected probabilities, $p_{\rm MC}=0.043$ and 0.048, but this is not a strong detection: the result does not approach a $3\sigma$ criterion, and is affected by the missing termination of one superoutburst. We therefore do not interpret the circular-statistics results as evidence for a physical orbital-phase trigger.

The absence of robust orbital-phase locking is physically plausible. In a thermonuclear nova, ignition occurs in the high-pressure layer on the WD surface, which should not retain direct information about the instantaneous azimuthal position of the donor star or accretion stream. For the active phases considered here, the relevant process is more likely the thermal-viscous state of the accretion disk. The timing of such disk-instability events should therefore be controlled mainly by the disk mass, radius, and thermal state, not by a unique binary phase. T CrB is not viewed face-on \citep{1998MNRAS.296...77B, 2004A&A...415..609S, 2025ApJ...983...76H}, so orbital geometry can in principle affect the observed accretion flow, but any such influence is expected to be indirect, for example through phase-dependent mass transfer, stream--disk impact geometry, or anisotropic red-giant wind structure. The essentially circular orbit further weakens any analogy with eccentric interacting binaries in which periastron passages can modulate the mass-transfer rate.

If the binary orbit directly triggered the active phases, one would expect a clear and robust concentration of onset phases and perhaps related clustering of maxima or terminations. Instead, the primary onset and maximum distributions are compatible with random sampling of orbital phase, and the termination distribution shows only a marginal hint. Any orbital influence must therefore be weaker than can be established robustly with the present data set and sample size.

\subsection{Relation to the SU UMa-like Interpretation}\label{subsec:suuma}

Our results are broadly consistent with the interpretation of \citet{2023ApJ...953L...7I}, who argued that the active phases of T CrB resemble normal outbursts and superoutbursts in SU~UMa-type dwarf novae. In that framework, the active phases are manifestations of accretion-disk instability rather than direct consequences of a specific orbital configuration. The lack of robust phase clustering, together with only a marginal termination hint, supports this general picture.

In the standard disk-instability and thermal--tidal-instability framework, normal outbursts are triggered by the thermal state of the accretion disk, while superoutbursts occur when the disk has expanded sufficiently for tidal instability, commonly associated with the 3:1 resonance in short-period SU~UMa systems \citep{1989PASJ...41.1005O, 2005PJAB...81..291O, 2013PASJ...65...50O}. Thus, the relevant ``phase'' in the usual SU~UMa phenomenology is primarily the supercycle phase, or the secular state of the disk within the normal-outburst/superoutburst cycle, rather than the instantaneous binary orbital phase. Empirical studies show that the orbital period is related to some global outburst properties, such as outburst duration, decline rate, and superhump period excess, and that the normal-outburst cycle length and superoutburst cycle length are correlated \citep{2016MNRAS.460.2526O}. These are statistical scaling relations across dwarf novae; they do not imply that individual onset, peak, or termination times should satisfy $t_{\rm event}=T_0+(N+\phi_0)P_{\rm orb}$.

At the same time, T CrB is not expected to behave like a standard short-period SU~UMa system in detail. In ordinary SU~UMa stars, the 3:1 tidal resonance plays a central role in producing superhumps and superoutburst behavior \citep{1988MNRAS.232...35W, 1989PASJ...41.1005O, 2013PASJ...65...23K, 2020PASJ...72...14K}. T CrB, however, is a symbiotic recurrent nova with a red-giant donor and a much larger mass ratio than the canonical SU~UMa regime. As emphasized by \citet{2023ApJ...953L...7I}, any analogy with SU~UMa stars must therefore be made cautiously, and the $\sim700$~d modulation reported during the most recent high state should not be interpreted as direct evidence that the standard 3:1 resonance mechanism operates unchanged in this system. The lack of robust orbital-phase clustering reinforces the view that T CrB is a long-period, symbiotic extension of dwarf-nova phenomenology rather than a simple scaled-up SU~UMa analog.

\subsection{Historical Nova Eruptions and the \citet{2024RNAAS...8..272S} Relation}\label{subsec:historical_discussion}

\citet{2024RNAAS...8..272S} pointed out that the separations between the four historical nova eruptions are approximately integer multiples of the orbital period. This is an intriguing empirical regularity, especially because the 1866 and 1946 eruptions are separated by nearly exactly 128 orbital cycles. However, our phase analysis does not establish a unique nova ignition phase. The two well-dated modern eruptions in 1866 and 1946 have very similar point-estimate phases, whereas the earlier 1217 and 1787 candidate eruptions lie at somewhat later phases under the same fixed modern ephemeris. This pattern is therefore suggestive at most, not evidence for one unique ignition phase, and its interpretation is further limited by the uncertainties of the early historical dates and the long-term period changes reported by \citet{2023MNRAS.524.3146S}.

The near coincidence between the 1866 and 1946 phases remains noteworthy, but it should not be over-interpreted. With only four historical eruptions, the statistical leverage is weak and post-hoc phase coincidences are not surprising. At present, the historical record is best viewed as motivation for continued monitoring rather than as evidence that nova ignition in T CrB is phase-locked to the orbit.

\subsection{The Secondary Eruption}\label{subsec:secondary}

The secondary eruption is one of the most remarkable aspects of T CrB. In both 1866 and 1946, a secondary brightening began after the primary nova had already faded and after the light curve had remained near its prior level for an extended interval \citep{2023MNRAS.524.3146S}. The two secondary eruptions begin at similar orbital phases, $\phi_{\rm sec,1866}\approx0.924$ and $\phi_{\rm sec,1946}\approx0.916$. This similarity should be recorded, but it should not be used as evidence for a physical phase-locking mechanism because the sample contains only two events.

A possible explanation is that the secondary brightening is connected to the recovery of accretion or to interaction between the nova ejecta and an asymmetric red-giant wind. Another possibility is the irradiation model of \citet{2023RNAAS...7..251M}, in which irradiation of the red giant by a cooling WD can reproduce the secondary maximum. In that scenario, similar secondary-maximum phases would not be independent of the geometry and timing of the preceding primary eruption. However, the present data do not distinguish between a phase-related mechanism, a post-eruption recovery timescale, or a coincidental similarity between the two best-observed eruptions. A future eruption and any subsequent secondary event would provide the decisive test.

\subsection{Implications}\label{subsec:implications}

The absence of robust phase locking places a useful constraint on models of T CrB. The active phases are unlikely to be triggered at a unique binary configuration. If orbital geometry contributes at all, its effect must be modest relative to the stochastic and secular evolution of the accretion disk, or it may appear only in specific operational descriptors such as termination boundaries. This conclusion also provides a conservative framework for interpreting the next eruption: future pre-eruption and post-eruption behavior should be monitored for phase-dependent signatures, but the current data do not justify predicting an eruption or active-phase boundary from orbital phase alone.

\section{Conclusions}\label{sec:conclusions}

We have tested whether the onset, maximum, and termination times of active phases in T CrB occur preferentially at specific orbital phases. Using long-term optical light curves, we measured characteristic times for four superoutbursts and ten normal outbursts, defining onset and termination by 15\% local-amplitude crossings in the processed $B$-band light curve. We converted these times to orbital phase with the circular spectroscopic ephemeris of \citet{2025ApJ...983...76H} and applied circular statistical tests.

Our conclusions are as follows.

\begin{enumerate}
\item The onset phases of all active phases are consistent with uniformity. The Kuiper and Watson probabilities for the full onset sample are $p_{\rm MC}=0.202$ and 0.190, respectively. We therefore treat the onset distribution as a non-detection of robust phase locking.

\item The orbital phases of outburst maxima are also consistent with a uniform distribution. The representative tests give $p_{\rm MC}=0.275$ and 0.345 for the Kuiper and Watson tests of all maxima, respectively. We find no evidence that maxima of either superoutbursts or normal outbursts are phase-locked to the orbit.

\item The 13 measurable termination phases give the lowest probabilities among the three primary timing definitions, with $p_{\rm MC}=0.043$ and 0.048 for the Kuiper and Watson tests, respectively. This is only marginal in an uncorrected sense, is far from a $3\sigma$ detection, and does not survive a simple correction for testing onset, maximum, and termination phases. We therefore regard the termination result as a cautionary hint rather than evidence for strict orbital-phase locking.

\item The four historical nova eruptions do not establish a unique orbital ignition phase. The phases are $0.5136$, $0.6028$, $0.4305$, and $0.4334$ for the 1217, 1787, 1866, and 1946 eruptions, respectively. Because the early dates are historically uncertain and the orbital period is not constant over long baselines, these phases should be interpreted only descriptively.

\item The secondary eruptions following the 1866 and 1946 novae occurred at similar fixed-ephemeris phases, but the sample size of two is too small to support a physical claim of orbital control. They also do not provide an independent phase-locking test if they are produced by post-nova irradiation of the red giant by the cooling white dwarf.
\end{enumerate}

Overall, the data do not support strict orbital-phase locking of the active phases in T CrB. The onset and maximum times are consistent with uniformity, while the termination times show at most a marginal hint. This conclusion is unchanged in the smoothing-radius sensitivity test described above. The onset, evolution, and termination of the outbursts are therefore more naturally interpreted as manifestations of intrinsic accretion-disk physics, with any orbital influence below the robust detection threshold of the present sample.

Future monitoring of T CrB remains essential. The next nova eruption and any subsequent secondary eruption will provide the best opportunity to test whether the historical phase coincidences are physically meaningful or coincidental.

\section*{Acknowledgements}
We extend our sincere gratitude to the anonymous referee for her or his insightful and constructive comments, which helped us to improve the scientific content of this article. We gratefully acknowledge the contributions of the AAVSO observer community, whose photometric data and metadata resources were used in this study and made available through the AAVSO's scientific archives. This work is supported by the High-level Talents Research Start-up Fund Project of Liupanshui Normal University (Grant No. LPSSYKYJJ202208), the Science and Technology Foundation of Guizhou Province (Grant Nos. QKHJC-ZK[2023]442 and QKHJCMS[2026]752), the Discipline-Team of Liupanshui Normal University (Grant No. LPSSY2023XKTD11), the Liupanshui Science and Technology Development Project (Grant Nos. 52020-2024-PT-01, 52020-2025-0-2-08, and 52020-2025-0-2-13), the National Natural Science Foundation of China (Grant Nos. 12303054, 12393853, and 12003020), the Yunnan Fundamental Research Projects (Grant Nos. 202401AU070063 and 202501AS070078), the National Key Research and Development Program of China (Grant No. 2024YFA1611603), the International Centre of Supernovae, Yunnan Key Laboratory (No. 202302AN360001), the High-Level Talent Recruitment Project of Qiannan Normal University for Nationalities (Grant No. qnsyrc202308), the Guizhou Provincial Basic Research Program (No. QKHJC[2024]youth158), and the Shaanxi Fundamental Science Research Project for Mathematics and Physics (Grant No. 23JSY015). R.Z.S. acknowledges support from the China Postdoctoral Science Foundation (Grant No. 2024M752979). Y.-Z. Cai acknowledges financial support from the SOXS project (PI S. Campana).

\facilities{AAVSO (AID)}

%% For this sample we use BibTeX plus aasjournalv7.bst to generate the

\bibliography{TCrB}{}

@ARTICLE{2023A&A...680L..18Z,
       author = {{Zamanov}, R. and {Boeva}, S. and {Latev}, G.~Y. and {Semkov}, E. and {Minev}, M. and {Kostov}, A. and {Bode}, M.~F. and {Marchev}, V. and {Marchev}, D.},
        title = "{Accretion in the recurrent nova T CrB: Linking the superactive state to the predicted outburst}",
      journal = {\aap},
         year = 2023,
        month = dec,
       volume = {680},
          eid = {L18},
        pages = {L18},
          doi = {10.1051/0004-6361/202348372},
archivePrefix = {arXiv},
       eprint = {2312.04342},
 primaryClass = {astro-ph.SR},
       adsurl = {https://ui.adsabs.harvard.edu/abs/2023A&A...680L..18Z}
}

@ARTICLE{1999A&AS..137..473M,
       author = {{M{\"u}rset}, U. and {Schmid}, H.~M.},
        title = "{Spectral classification of the cool giants in symbiotic systems}",
      journal = {\aaps},
         year = 1999,
        month = jun,
       volume = {137},
        pages = {473-493},
          doi = {10.1051/aas:1999105},
       adsurl = {https://ui.adsabs.harvard.edu/abs/1999A&AS..137..473M}
}

@ARTICLE{1998MNRAS.296...77B,
       author = {{Belczynski}, K. and {Mikolajewska}, J.},
        title = "{New binary parameters for the symbiotic recurrent nova T Coronae Borealis}",
      journal = {\mnras},
         year = 1998,
        month = may,
       volume = {296},
       number = {1},
        pages = {77-84},
          doi = {10.1046/j.1365-8711.1998.01301.x},
archivePrefix = {arXiv},
       eprint = {astro-ph/9711151},
 primaryClass = {astro-ph},
       adsurl = {https://ui.adsabs.harvard.edu/abs/1998MNRAS.296...77B}
}

@ARTICLE{2004A&A...415..609S,
       author = {{Stanishev}, V. and {Zamanov}, R. and {Tomov}, N. and {Marziani}, P.},
        title = "{H{\ensuremath{\alpha}} variability of the recurrent nova T Coronae Borealis}",
      journal = {\aap},
         year = 2004,
        month = feb,
       volume = {415},
        pages = {609-616},
          doi = {10.1051/0004-6361:20034623},
archivePrefix = {arXiv},
       eprint = {astro-ph/0311309},
 primaryClass = {astro-ph},
       adsurl = {https://ui.adsabs.harvard.edu/abs/2004A&A...415..609S}
}

@ARTICLE{1949ApJ...109...81S,
       author = {{Sanford}, Roscoe F.},
        title = "{High-Dispersion Spectrograms of T Coronae Borealis.}",
      journal = {\apj},
         year = 1949,
        month = jan,
       volume = {109},
        pages = {81},
          doi = {10.1086/145106},
       adsurl = {https://ui.adsabs.harvard.edu/abs/1949ApJ...109...81S}
}

@ARTICLE{2023JHA....54..436S,
       author = {{Schaefer}, Bradley E.},
        title = "{The recurrent nova T CrB had prior eruptions observed near December 1787 and October 1217 AD}",
      journal = {Journal for the History of Astronomy},
         year = 2023,
        month = nov,
       volume = {54},
       number = {4},
        pages = {436-455},
          doi = {10.1177/00218286231200492},
archivePrefix = {arXiv},
       eprint = {2308.13668},
 primaryClass = {astro-ph.SR},
       adsurl = {https://ui.adsabs.harvard.edu/abs/2023JHA....54..436S}
}

@ARTICLE{2020ApJ...902L..14L,
       author = {{Luna}, Gerardo J.~M. and {Sokoloski}, J.~L. and {Mukai}, Koji and {M. Kuin}, N. Paul},
        title = "{Increasing Activity in T CrB Suggests Nova Eruption Is Impending}",
      journal = {\apjl},
         year = 2020,
        month = oct,
       volume = {902},
       number = {1},
          eid = {L14},
        pages = {L14},
          doi = {10.3847/2041-8213/abbb2c},
archivePrefix = {arXiv},
       eprint = {2009.11902},
 primaryClass = {astro-ph.SR},
       adsurl = {https://ui.adsabs.harvard.edu/abs/2020ApJ...902L..14L}
}

@ARTICLE{2023ATel16107....1S,
       author = {{Schaefer}, Bradley E. and {Kloppenborg}, Brian and {Waagen}, Elizabeth O. and {Observers}, The Aavso},
        title = "{Recurrent nova T CrB has just started its Pre-eruption Dip in March/April 2023, so the eruption should occur around 2024.4 +- 0.3}",
      journal = {The Astronomer's Telegram},
         year = 2023,
        month = jun,
       volume = {16107},
        pages = {1},
       adsurl = {https://ui.adsabs.harvard.edu/abs/2023ATel16107....1S}
}

@ARTICLE{2016NewA...47....7M,
       author = {{Munari}, Ulisse and {Dallaporta}, Sergio and {Cherini}, Giulio},
        title = "{The 2015 super-active state of recurrent nova T CrB and the long term evolution after the 1946 outburst}",
      journal = {\na},
         year = 2016,
        month = aug,
       volume = {47},
        pages = {7-15},
          doi = {10.1016/j.newast.2016.01.002},
archivePrefix = {arXiv},
       eprint = {1602.07470},
 primaryClass = {astro-ph.SR},
       adsurl = {https://ui.adsabs.harvard.edu/abs/2016NewA...47....7M}
}

@ARTICLE{2016MNRAS.462.2695I,
       author = {{I{\l}kiewicz}, Krystian and {Miko{\l}ajewska}, Joanna and {Stoyanov}, Kiril and {Manousakis}, Antonios and {Miszalski}, Brent},
        title = "{Active phases and flickering of a symbiotic recurrent nova T CrB}",
      journal = {\mnras},
         year = 2016,
        month = nov,
       volume = {462},
       number = {3},
        pages = {2695-2705},
          doi = {10.1093/mnras/stw1837},
archivePrefix = {arXiv},
       eprint = {1607.06804},
 primaryClass = {astro-ph.SR},
       adsurl = {https://ui.adsabs.harvard.edu/abs/2016MNRAS.462.2695I}
}

@ARTICLE{2019ApJ...884....8L,
       author = {{Linford}, Justin D. and {Chomiuk}, Laura and {Sokoloski}, Jennifer L. and {Weston}, Jennifer H.~S. and {van der Horst}, Alexander J. and {Mukai}, Koji and {Barrett}, Paul and {Mioduszewski}, Amy J. and {Rupen}, Michael},
        title = "{T CrB: Radio Observations during the 2016-2017 {\textquotedblleft}Super-active{\textquotedblright} State}",
      journal = {\apj},
         year = 2019,
        month = oct,
       volume = {884},
       number = {1},
          eid = {8},
        pages = {8},
          doi = {10.3847/1538-4357/ab3c62},
archivePrefix = {arXiv},
       eprint = {1909.13858},
 primaryClass = {astro-ph.HE},
       adsurl = {https://ui.adsabs.harvard.edu/abs/2019ApJ...884....8L}
}

@ARTICLE{2023ApJ...953L...7I,
       author = {{I{\l}kiewicz}, Krystian and {Miko{\l}ajewska}, Joanna and {Stoyanov}, Kiril A.},
        title = "{Symbiotic Star T CrB as an Extreme SU UMa-type Dwarf Nova}",
      journal = {\apjl},
         year = 2023,
        month = aug,
       volume = {953},
       number = {1},
          eid = {L7},
        pages = {L7},
          doi = {10.3847/2041-8213/ace9dc},
archivePrefix = {arXiv},
       eprint = {2307.13838},
 primaryClass = {astro-ph.SR},
       adsurl = {https://ui.adsabs.harvard.edu/abs/2023ApJ...953L...7I}
}

@ARTICLE{2025MNRAS.541L..14M,
       author = {{Merc}, Jaroslav and {Wyrzykowski}, {\L}ukasz and {Beck}, Paul G. and {Miko{\l}ajczyk}, Przemys{\l}aw J. and {Kotysz}, Krzysztof and {Zieli{\'n}ski}, Pawe{\l} and {Zola}, Staszek and {Kurowski}, Sebastian and {Og{\l}oza}, Waldemar and {Drozdz}, Marek and {Galdies}, Charles and {Hambsch}, Franz-Josef and {Brincat}, Stephen M. and {Joachimczyk}, Barbara and {Bronikowski}, Mateusz and {Japelj}, Jure and {Mihelcic}, Matej and {Carrasco}, Josep Manel and {Burgaz}, Umut and {Gurgul}, Agnieszka and {B{\k{a}}kowska}, Karolina and {Hofbauer}, Piotr and {Szyszka}, Krzysztof and {Golonka}, Jan and {Qvam}, Jan K{\r{a}}re Trandem and {Zdanavi{\v{c}}ius}, Justas and {Pak{\v{s}}tien{\.{e}}}, Erika and {Maskoli{\={u}}nas}, Marius and {{\v{C}}epas}, Vytautas and {Pylypenko}, Uliana and {Mo{\'z}dzierski}, Dawid and {Dubois}, Franky and {Vanaverbeke}, Siegfried and {Olszewska}, Justyna M. and {Liakos}, Alexios and {Stojanovi{\'c}}, Milan and {Damljanovi{\'c}}, Goran and {Popowicz}, Adam and {Marzec}, Mateusz and {Badura}, Magdalena and {Gil}, Bartosz and {Pucek}, Alicja and {Kowalska}, Aleksandra and {Szklarz}, Mateusz and {Kvernadze}, Teimuraz and {Reguitti}, Andrea and {Awiphan}, Supachai and {Dennefeld}, Michel and {Gazeas}, Kosmas},
        title = "{Is the symbiotic recurrent nova T CrB late? Recent photometric evolution and comparison with past pre-outburst behaviour}",
      journal = {\mnras},
         year = 2025,
        month = jul,
       volume = {541},
       number = {1},
        pages = {L14-L21},
          doi = {10.1093/mnrasl/slaf047},
archivePrefix = {arXiv},
       eprint = {2504.20592},
 primaryClass = {astro-ph.SR},
       adsurl = {https://ui.adsabs.harvard.edu/abs/2025MNRAS.541L..14M}
}

@ARTICLE{2025A&A...694A..85P,
       author = {{Planquart}, L. and {Jorissen}, A. and {Van Winckel}, H.},
        title = "{Resolving the mass transfer in the symbiotic recurrent nova T Coronae Borealis}",
      journal = {\aap},
         year = 2025,
        month = feb,
       volume = {694},
          eid = {A85},
        pages = {A85},
          doi = {10.1051/0004-6361/202452833},
archivePrefix = {arXiv},
       eprint = {2501.02984},
 primaryClass = {astro-ph.SR},
       adsurl = {https://ui.adsabs.harvard.edu/abs/2025A&A...694A..85P}
}

@ARTICLE{2018A&A...619A..61L,
       author = {{Luna}, G.~J.~M. and {Mukai}, K. and {Sokoloski}, J.~L. and {Nelson}, T. and {Kuin}, P. and {Segreto}, A. and {Cusumano}, G. and {Jaque Arancibia}, M. and {Nu{\~n}ez}, N.~E.},
        title = "{Dramatic change in the boundary layer in the symbiotic recurrent nova T Coronae Borealis}",
      journal = {\aap},
         year = 2018,
        month = nov,
       volume = {619},
          eid = {A61},
        pages = {A61},
          doi = {10.1051/0004-6361/201833747},
archivePrefix = {arXiv},
       eprint = {1807.01304},
 primaryClass = {astro-ph.HE},
       adsurl = {https://ui.adsabs.harvard.edu/abs/2018A&A...619A..61L}
}

@ARTICLE{2025ApJ...983...76H,
       author = {{Hinkle}, Kenneth H. and {Nagarajan}, Pranav and {Fekel}, Francis C. and {Miko{\l}ajewska}, Joanna and {Straniero}, Oscar and {Muterspaugh}, Matthew W.},
        title = "{Binary Parameters for the Recurrent Nova T Coronae Borealis}",
      journal = {\apj},
         year = 2025,
        month = apr,
       volume = {983},
       number = {1},
          eid = {76},
        pages = {76},
          doi = {10.3847/1538-4357/adbe63},
archivePrefix = {arXiv},
       eprint = {2502.20664},
 primaryClass = {astro-ph.SR},
       adsurl = {https://ui.adsabs.harvard.edu/abs/2025ApJ...983...76H}
}

@ARTICLE{1992ApJ...393..289S,
       author = {{Selvelli}, Pier L. and {Cassatella}, Angelo and {Gilmozzi}, Roberto},
        title = "{The Nature of the Recurrent Nova T Coronae Borealis: Ultraviolet Evidence for a White Dwarf Accretor}",
      journal = {\apj},
         year = 1992,
        month = jul,
       volume = {393},
        pages = {289},
          doi = {10.1086/171506},
       adsurl = {https://ui.adsabs.harvard.edu/abs/1992ApJ...393..289S}
}

@ARTICLE{1975JBAA...85..217B,
       author = {{Bailey}, J.},
        title = "{Periodic fluctuations in the recurrent nova T CrB.}",
      journal = {Journal of the British Astronomical Association},
         year = 1975,
        month = mar,
       volume = {85},
        pages = {217-223},
       adsurl = {https://ui.adsabs.harvard.edu/abs/1975JBAA...85..217B}
}

@ARTICLE{2000AJ....119.1375F,
       author = {{Fekel}, Francis C. and {Joyce}, Richard R. and {Hinkle}, Kenneth H. and {Skrutskie}, Michael F.},
        title = "{Infrared Spectroscopy of Symbiotic Stars. I. Orbits for Well-Known S-Type Systems}",
      journal = {\aj},
         year = 2000,
        month = mar,
       volume = {119},
       number = {3},
        pages = {1375-1388},
          doi = {10.1086/301260},
       adsurl = {https://ui.adsabs.harvard.edu/abs/2000AJ....119.1375F}
}

@ARTICLE{2025A&A...701A.176M,
       author = {{Munari}, U. and {Walter}, F. and {Masetti}, N. and {Valisa}, P. and {Dallaporta}, S. and {Bergamini}, A. and {Cherini}, G. and {Frigo}, A. and {Maitan}, A. and {Marino}, C. and {Mazzacurati}, G. and {Moretti}, S. and {Tabacco}, F. and {Tomaselli}, S. and {Vagnozzi}, A. and {Ochner}, P. and {Albanese}, I.},
        title = "{T CrB: Overview of the accretion history, Roche-lobe filling, orbital solution, and radiative modeling}",
      journal = {\aap},
         year = 2025,
        month = sep,
       volume = {701},
          eid = {A176},
        pages = {A176},
          doi = {10.1051/0004-6361/202555917},
       adsurl = {https://ui.adsabs.harvard.edu/abs/2025A&A...701A.176M}
}

@ARTICLE{2004MNRAS.350.1477Z,
       author = {{Zamanov}, R. and {Bode}, M.~F. and {Stanishev}, V. and {Mart{\'\i}}, J.},
        title = "{Flickering variability of T Coronae Borealis}",
      journal = {\mnras},
         year = 2004,
        month = jun,
       volume = {350},
       number = {4},
        pages = {1477-1484},
          doi = {10.1111/j.1365-2966.2004.07747.x},
archivePrefix = {arXiv},
       eprint = {astro-ph/0402465},
 primaryClass = {astro-ph},
       adsurl = {https://ui.adsabs.harvard.edu/abs/2004MNRAS.350.1477Z}
}

@BOOK{1992msp..book.....S,
       author = {{Strai{\v{z}}ys}, Vytautas},
        title = "{Multicolor stellar photometry}",
         year = 1992,
       adsurl = {https://ui.adsabs.harvard.edu/abs/1992msp..book.....S}
}

@ARTICLE{2024RNAAS...8..272S,
       author = {{Schneider}, Jean},
        title = "{When will the Next T CrB Eruption Occur?}",
      journal = {Research Notes of the American Astronomical Society},
         year = 2024,
        month = oct,
       volume = {8},
       number = {10},
          eid = {272},
        pages = {272},
          doi = {10.3847/2515-5172/ad8bba},
       adsurl = {https://ui.adsabs.harvard.edu/abs/2024RNAAS...8..272S}
}

@INPROCEEDINGS{1997ppsb.conf..117A,
       author = {{Anupama}, G.~C.},
        title = "{The Optical Emission Line Variability in T CrB}",
    booktitle = {Physical Processes in Symbiotic Binaries and Related Systems},
         year = 1997,
       editor = {{Miko{\l}ajewska}, Joanna},
        month = jan,
        pages = {117},
       adsurl = {https://ui.adsabs.harvard.edu/abs/1997ppsb.conf..117A}
}

@ARTICLE{2026A&A...706A..94P,
       author = {{Pei}, Songpeng and {Zhang}, Xiaowan and {Su}, Renzhi and {Cai}, Yongzhi and {Ou}, Ziwei and {Li}, Qiang and {Ren}, Xiaoqin and {Yang}, Taozhi and {Li}, Mingyue},
        title = "{Multiwavelength study of the pre-eruption dip in the recurrent nova T Coronae Borealis preceding imminent nova eruption}",
      journal = {\aap},
         year = 2026,
        month = feb,
       volume = {706},
          eid = {A94},
        pages = {A94},
          doi = {10.1051/0004-6361/202557346},
archivePrefix = {arXiv},
       eprint = {2512.19218},
 primaryClass = {astro-ph.HE},
       adsurl = {https://ui.adsabs.harvard.edu/abs/2026A&A...706A..94P}
}

@ARTICLE{2023MNRAS.524.3146S,
       author = {{Schaefer}, Bradley E.},
        title = "{The B \& V light curves for recurrent nova T CrB from 1842-2022, the unique pre- and post-eruption high-states, the complex period changes, and the upcoming eruption in 2025.5 {\ensuremath{\pm}} 1.3}",
      journal = {\mnras},
         year = 2023,
        month = sep,
       volume = {524},
       number = {2},
        pages = {3146-3165},
          doi = {10.1093/mnras/stad735},
archivePrefix = {arXiv},
       eprint = {2303.04933},
 primaryClass = {astro-ph.SR},
       adsurl = {https://ui.adsabs.harvard.edu/abs/2023MNRAS.524.3146S}
}

@ARTICLE{1988MNRAS.232...35W,
       author = {{Whitehurst}, Robert},
        title = "{Numerical simulations of accretion discs - I. Superhumps : a tidal phenomenon of accretion discs.}",
      journal = {\mnras},
         year = 1988,
        month = may,
       volume = {232},
        pages = {35-51},
          doi = {10.1093/mnras/232.1.35},
       adsurl = {https://ui.adsabs.harvard.edu/abs/1988MNRAS.232...35W}
}

@ARTICLE{1989PASJ...41.1005O,
       author = {{Osaki}, Yoji},
        title = "{A Model for the Superoutburst Phenomenon of SU Ursae Majoris Stars}",
      journal = {\pasj},
         year = 1989,
        month = dec,
       volume = {41},
       number = {5},
        pages = {1005-1033},
          doi = {10.1093/pasj/41.5.1005},
       adsurl = {https://ui.adsabs.harvard.edu/abs/1989PASJ...41.1005O}
}

@ARTICLE{2013PASJ...65...23K,
       author = {{Kato}, Taichi and {Hambsch}, Franz-Josef and {Maehara}, Hiroyuki and {Masi}, Gianluca and {Miller}, Ian and {Noguchi}, Ryo and {Akasaka}, Chihiro and {Aoki}, Tomoya and {Kobayashi}, Hiroshi and {Matsumoto}, Katsura and {Nakagawa}, Shinichi and {Nakazato}, Takuma and {Nomoto}, Takashi and {Ogura}, Kazuyuki and {Ono}, Rikako and {Taniuchi}, Keisuke and {Stein}, William and {Henden}, Arne and {de Miguel}, Seiichiro, Enrique Kiyota and {Dubovsky}, Pavol A. and {Kudzej}, Igor and {Imamura}, Kazuyoshi and {Akazawa}, Hidehiko and {Takagi}, Ryosuke and {Wakabayashi}, Yuya and {Ogi}, Minako and {Tanabe}, Kenji and {Ulowetz}, Joseph and {Morelle}, Etienne and {Pickard}, Roger D. and {Ohshima}, Tomohito and {Kasai}, Kiyoshi and {Pavlenko}, Elena P. and {Antonyuk}, Oksana I. and {Baklanov}, Aleksei V. and {Antonyuk}, Kirill and {Samsonov}, Denis and {Pit}, Nikolaj and {Sosnovskij}, Aleksei and {Littlefield}, Colin and {Sabo}, Richard and {Ruiz}, Javier and {Krajci}, Thomas and {Dvorak}, Shawn and {Oksanen}, Arto and {Hirosawa}, Kenji and {Goff}, William N. and {Monard}, Berto and {Shears}, Jeremy and {Boyd}, David and {Voloshina}, Irina B. and {Shugarov}, Sergey Yu. and {Chochol}, Drahomir and {Miyashita}, Atsushi and {Pietz}, Jochen and {Katysheva}, Natalia and {Itoh}, Hiroshi and {Bolt}, Greg and {Andreev}, Maksim V. and {Parakhin}, Nikolai and {Malanushenko}, Viktor and {Martinelli}, Fabio and {Denisenko}, Denis and {Stockdale}, Chris and {Starr}, Peter and {Simonsen}, Mike and {Tristram}, Paul J. and {Fukui}, Akihiko and {Tordai}, Tamas and {Fidrich}, Robert and {Paxson}, Kevin B. and {Itagaki}, Koh-ichi and {Nakashima}, Youichirou and {Yoshida}, Seiichi and {Nishimura}, Hideo and {Kryachko}, Timur V. and {Samokhvalov}, Andrey V. and {Korotkiy}, Stanislav A. and {Satovski}, Boris L. and {Stubbings}, Rod and {Poyner}, Gary and {Muyllaert}, Eddy and {Gerke}, Vladimir and {MacDonald}, II, Walter and {Linnolt}, Michael and {Maeda}, Yutaka and {Hautecler}, Hubert},
        title = "{Survey of Period Variations of Superhumps in SU UMa-Type Dwarf Novae. IV. The Fourth Year (2011-2012)}",
      journal = {\pasj},
         year = 2013,
        month = feb,
       volume = {65},
          eid = {23},
        pages = {23},
          doi = {10.1093/pasj/65.1.23},
archivePrefix = {arXiv},
       eprint = {1210.0678},
 primaryClass = {astro-ph.SR},
       adsurl = {https://ui.adsabs.harvard.edu/abs/2013PASJ...65...23K}
}

@ARTICLE{2020PASJ...72...14K,
       author = {{Kato}, Taichi and {Isogai}, Keisuke and {Wakamatsu}, Yasuyuki and {Hambsch}, Franz-Josef and {Itoh}, Hiroshi and {Tordai}, Tam{\'a}s and {Vanmunster}, Tonny and {Dubovsky}, Pavol A. and {Kudzej}, Igor and {Medulka}, Tom{\'a}{\v{s}} and {Kimura}, Mariko and {Ohnishi}, Ryuhei and {Monard}, Berto and {Pavlenko}, Elena P. and {Antonyuk}, Kirill A. and {Pit}, Nikolaj V. and {Antonyuk}, Oksana I. and {Babina}, Julia V. and {Baklanov}, Aleksei V. and {Sosnovskij}, Aleksei A. and {Pickard}, Roger D. and {Miller}, Ian and {Maeda}, Yutaka and {de Miguel}, Enrique and {Brincat}, Stephen M. and {Licchelli}, Domenico and {Cook}, Lewis M. and {Shugarov}, Sergey Yu and {Zaostrojnykh}, Anna M. and {Chochol}, Drahomir and {Golysheva}, Polina and {Katysheva}, Natalia and {Zubareva}, Alexandra M. and {Stone}, Geoff and {Kasai}, Kiyoshi and {Starr}, Peter and {Littlefield}, Colin and {Kiyota}, Seiichiro and {Andreev}, Maksim V. and {Sergeev}, Alexandr V. and {Ruiz}, Javier and {Myers}, Gordon and {Simon}, Andrii O. and {Vasylenko}, Volodymyr V. and {Sold{\'a}n}, Francisco and {{\"O}gmen}, Yenal and {Nakajima}, Kazuhiro and {Nelson}, Peter and {Masi}, Gianluca and {Menzies}, Kenneth and {Sabo}, Richard and {Bolt}, Greg and {Dvorak}, Shawn and {Stanek}, Krzysztof Z. and {Shields}, Joseph V. and {Kochanek}, Christopher S. and {Holoien}, Thomas W.-S. and {Shappee}, Benjamin and {Prieto}, Jos{\'e} L. and {Kojima}, Tadashi and {Nishimura}, Hideo and {Kaneko}, Shizuo and {Fujikawa}, Shigehisa and {Stubbings}, Rod and {Muyllaert}, Eddy and {Poyner}, Gary and {Moriyama}, Masayuki and {Maehara}, Hiroyuki and {Schmeer}, Patrick and {Denisenko}, Denis},
        title = "{Survey of period variations of superhumps in SU UMa-type dwarf novae. X. The tenth year (2017)}",
      journal = {\pasj},
         year = 2020,
        month = feb,
       volume = {72},
       number = {1},
          eid = {14},
        pages = {14},
          doi = {10.1093/pasj/psz134},
archivePrefix = {arXiv},
       eprint = {1911.04645},
 primaryClass = {astro-ph.SR},
       adsurl = {https://ui.adsabs.harvard.edu/abs/2020PASJ...72...14K}
}

@misc{2022zndo...6344451M,
       author = {{Marholm}, Sigvald},
        title = "{sigvaldm/localreg: Multivariate RBF output, v0.5.0, Zenodo, doi:10.5281/zenodo.6344451}",
        journal = {\Zenodo},
         year = 2022,
        month = mar,
        volume = {doi:10.5281/zenodo.6344451},
          eid = {10.5281/zenodo.6344451},
          doi = {10.5281/zenodo.6344451},
      version = {0.5.0},
    publisher = {Zenodo},
       adsurl = {https://ui.adsabs.harvard.edu/abs/2022zndo...6344451M}
}

@ARTICLE{1997IBVS.4461....1Z,
       author = {{Zamanov}, R.~K. and {Zamanova}, V.~I.},
        title = "{UBV Observations of T CrB}",
      journal = {Information Bulletin on Variable Stars},
         year = 1997,
        month = mar,
       volume = {4461},
        pages = {1},
       adsurl = {https://ui.adsabs.harvard.edu/abs/1997IBVS.4461....1Z}
}

@ARTICLE{Kuiper1960,
  author  = {Kuiper, N. H.},
  title   = {Tests concerning random points on a circle},
  journal = {Proceedings of the Koninklijke Nederlandse Akademie van Wetenschappen, Series A},
  volume  = {63},
  pages   = {38--47},
  year    = {1960}
}

@ARTICLE{Watson1961,
  author  = {Watson, G. S.},
  title   = {Goodness-of-fit tests on a circle},
  journal = {Biometrika},
  volume  = {48},
  number  = {1--2},
  pages   = {109--114},
  year    = {1961},
  doi     = {10.1093/biomet/48.1-2.109}
}

@BOOK{Press2007,
  author    = {Press, William H. and Teukolsky, Saul A. and Vetterling, William T. and Flannery, Brian P.},
  title     = {Numerical Recipes: The Art of Scientific Computing},
  edition   = {3},
  publisher = {Cambridge University Press},
  address   = {Cambridge},
  year      = {2007}
}

@BOOK{Jammalamadaka2001,
  author    = {Jammalamadaka, S. Rao and SenGupta, A.},
  title     = {Topics in Circular Statistics},
  publisher = {World Scientific},
  address   = {Singapore},
  year      = {2001}
}

@ARTICLE{2005PJAB...81..291O,
       author = {{Osaki}, Yoji},
        title = "{The disk instability model for dwarf nova outbursts}",
      journal = {Proceedings of the Japan Academy, Series B},
         year = 2005,
        month = nov,
       volume = {81},
        pages = {291-305},
          doi = {10.2183/pjab.81.291},
       adsurl = {https://ui.adsabs.harvard.edu/abs/2005PJAB...81..291O}
}

@ARTICLE{2013PASJ...65...50O,
       author = {{Osaki}, Yoji and {Kato}, Taichi},
        title = "{The Cause of the Superoutburst in SU UMa Stars is Finally Revealed by Kepler Light Curve of V1504 Cygni}",
      journal = {\pasj},
         year = 2013,
        month = jun,
       volume = {65},
          eid = {50},
        pages = {50},
          doi = {10.1093/pasj/65.3.50},
archivePrefix = {arXiv},
       eprint = {1212.1516},
 primaryClass = {astro-ph.SR},
       adsurl = {https://ui.adsabs.harvard.edu/abs/2013PASJ...65...50O}
}

@ARTICLE{2016MNRAS.460.2526O,
       author = {{Otulakowska-Hypka}, Magdalena and {Olech}, Arkadiusz and {Patterson}, Joseph},
        title = "{Statistical analysis of properties of dwarf novae outbursts}",
      journal = {\mnras},
         year = 2016,
        month = aug,
       volume = {460},
       number = {3},
        pages = {2526-2541},
          doi = {10.1093/mnras/stw1120},
archivePrefix = {arXiv},
       eprint = {1605.02937},
 primaryClass = {astro-ph.SR},
       adsurl = {https://ui.adsabs.harvard.edu/abs/2016MNRAS.460.2526O}
}

@misc{Kloppenborg2025AAVSO,
  author       = {Kloppenborg, B. K.},
  year         = {2025},
  title        = {{Observations from the AAVSO International Database}},
  howpublished = {\url{https://www.aavso.org}},
  note         = {Accessed for T CrB photometry}
}

@ARTICLE{2023RNAAS...7..251M,
       author = {{Munari}, Ulisse},
        title = "{The Secondary Maximum of T CrB Caused by Irradiation of the Red Giant by a Cooling White Dwarf}",
      journal = {Research Notes of the American Astronomical Society},
         year = 2023,
        month = nov,
       volume = {7},
       number = {11},
          eid = {251},
        pages = {251},
          doi = {10.3847/2515-5172/ad0f26},
archivePrefix = {arXiv},
       eprint = {2311.15909},
 primaryClass = {astro-ph.SR},
       adsurl = {https://ui.adsabs.harvard.edu/abs/2023RNAAS...7..251M}
}
\bibliographystyle{aasjournalv7}



\end{document}